\documentclass[onecolumn,10pt]{article}

\usepackage{amsmath}

\usepackage{amssymb}

\usepackage{hyperref}

\usepackage{graphicx}

\usepackage{cite}

\usepackage{graphics,dcolumn,bm,epic,eepic,float}

\usepackage{epstopdf} 

\usepackage{authblk} 

\usepackage{verbatim}

\usepackage{xcolor,colortbl}

\usepackage[labelfont=bf,labelsep=period,justification=raggedright,small]{caption}

\usepackage{subcaption}

\date{}

\begin{document}

\title{The Impact of Social Segregation on Human Mobility in Developing and Urbanized Regions}

\author[]{Alexander Amini}
\author[]{Kevin Kung}
\author[]{Chaogui Kang}
\author[]{Stanislav Sobolevsky}
\author[]{Carlo Ratti}
\affil[]{SENSEable City Laboratory, Massachusetts Institute of Technology, Cambridge, MA 02139, USA}
\maketitle

\begin{abstract}

This study leverages mobile phone data to analyze human mobility patterns in developing countries, especially in comparison to more industrialized countries.   Developing regions, such as the Ivory Coast, are marked by a number of factors that may influence mobility, such as less infrastructural coverage and maturity, less economic resources and stability, and in some cases, more cultural and language-based diversity.  By comparing mobile phone data collected from the Ivory Coast to similar data collected in Portugal, we are able to highlight both qualitative and quantitative differences in mobility patterns - such as differences in likelihood to travel, as well as in the time required to travel - that are relevant to consideration on policy, infrastructure, and economic development. Our study illustrates how cultural and linguistic diversity in developing regions (such as Ivory Coast) can present challenges to mobility models that perform well and were conceptualized in less culturally diverse regions. Finally, we address these challenges by proposing novel techniques to assess the strength of borders in a regional partitioning scheme and to quantify the impact of border strength on mobility model accuracy.
\end{abstract}

\section{Introduction}
Transportation and communication networks form the fabric of industrialized nations.  The rollout of such infrastructure in such regions can play a major role in supporting, or deterring, a region’s ability to thrive economically and socially. Likewise, citizens’ use of these networks can tell us much about the region, including insight on how ideas and diseases may be spreading, or how to most effectively augment services, such as healthcare and education \cite{robertson2010disease}.

Existing studies of mobile phone data have given us insight on numerous aspects of human mobility \cite{gonzalez2008mobility, ratti2006mobile, reades2009eigenplaces, simini2012radiation, kang2013exploring}. However, these studies tend to focus on regions with the highest mobile phone coverage, which also happens to be in more stable, mature, and developed regions. Thus, the models produced based on this data might not be as appropriate for developing regions with a substantially different patterns of social interactions and human mobility. However, these highly industrialized and wealthy regions represent less than one-third of the world’s population, with the remaining two-thirds living in developing and less economically mature regions. Although it is the developing regions which are facing the most rapid demographic and economic shifts worldwide, and are in even greater need of such models to help inform policy makers, urban planners, and service providers. Yet, little work has been done to assess the appropriateness of models conceptualized for industrialized regions for use in developing regions.

Obtaining a comprehensive and accurate dataset of the telecommunication activity in developing regions can be extremely difficult for a number of reasons; however, the Data4Developoment (D4D) dataset \cite{blondel2012d4d} provided a unique opportunity by collecting data throughout the Ivory Coast and releasing it specifically for research purposes, so that developing regions could also be analyzed in greater detail. With over 60 distinct tribes \cite{ic_online}, Ivory Coast boasts rich cultural and linguistic diversity, in addition to its rapid urbanization. These contrasting social interactions offer researchers a unique opportunity to understand the communication and mobility patterns and needs of a developing nation during key phases of its transformation. 

Our goal was to leverage the contrast between mobility data from Ivory Coast with a more industrialized nation (Portugal) in order to assess the ability of human mobility models developed for industrialized regions to accurately model developing regions.

Our findings shed new light on the applicability of metrics and models conceptualized for industrialized regions, to developing regions. Our results demonstrate the importance of considering cultural and linguistic diversity in the construction of new models to address the challenges of developing regions. The insights gained from our study have important applications to policymaking, urban planning, and the services deployments that are transforming Ivory Coast and many other developing countries.

In the following sections, we provide additional details on the data used in this study, the results derived, and the conclusions drawn.

\section{Data Description}

We used five datasets to assess and compare the human mobility patterns in the Ivory Coast and Portugal. The first two datasets, \textit{D1} and \textit{D2}, were provided by Orange telecom as \textit{SET1} and \textit{SET2} respectively, via the Data for Development (D4D) Challenge \cite{blondel2012d4d}. Both datasets are based on anonymized Call Detail Records (CDRs) of 2.5 billion calls and SMS exchanges between 5 million users December 1, 2011 until April 28, 2012 (150 days). 

\textit{SET1} provided the number and duration of all calls between any pair of antenna, aggregated by hour. Calls spanning multiple time slots were considered to be in the time slot when they began. Communication between Orange customers and customers of other providers were removed.

\begin{figure}[b!]
	\begin{center}
		\includegraphics[keepaspectratio=false, width=10.3cm, height=15.0cm, trim=3.4cm 4.6cm 1.4cm 3cm]{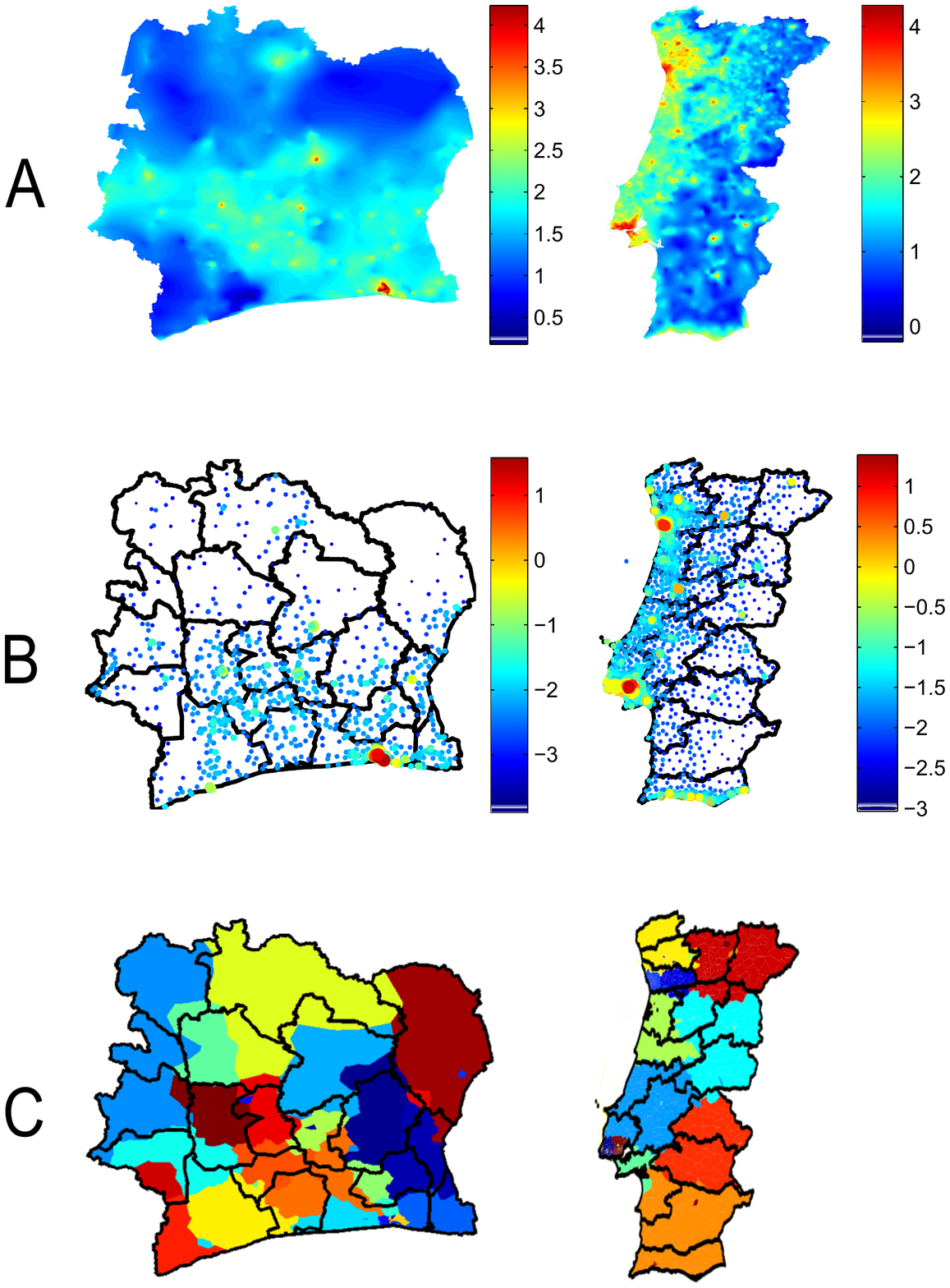}
	\end{center}   
	\caption{Cartographic representations of the Ivory Coast (left) and Portugal. \textit{A. Population Distribution}, where colors are logarithmically based on the population density. \textit{B. Geographic cell tower position}, where the size and color of the antennae are logarithmically mapped based on the density of antennae in the area. \textit{C. Community Partitioning}, where communities are built from human migrations over a 24 hour time window and visually displayed along with the countries official administrative boundaries.}
\label{intro_maps}
\end{figure}

\textit{SET2} contains consecutive call activities of each subscriber over the study period. Each record in this dataset represents a single connection to an antenna and contains the following fields: timestamp, anonymized ID of the user, and the antenna ID they connected to. To further anonymize this data, the original dataset was subsampled to the calls of 50K randomly sampled individuals for each of 2-week periods in the dataset. The geographical positions of the antenna for \textit{D1} and \textit{D2} were also provided and are illustrated in Figure \ref{intro_maps}B. Records without antenna IDs were removed; 107 antennae had no calls and 128 antennae had no population movements. 

The third dataset, \textit{D3}, was also provided by Orange with over 2000 antennae distributed across Portugal. \textit{D3} included the same fields as \textit{D2}, for 400 million anonymized CDRs from 2 million users, for the time period of January 1, 2006, to December 31, 2007.

Datasets \textit{D4} and \textit{D5} provided a high-resolution population density data for Ivory Coast \cite{afripop} and Portugal \cite{portugal_stats}, respectively. To map the population data to the antennae, we created a Voronoi tessellation \cite{voronoi1908} of each country based on the antennae location. For the 12 locations that had 2-3 antennas in a single location, those 2-3 antennas were collapsed into a single Voronoi cell. Each antenna was assigned the total population within the corresponding Voronoi cell. Figure \ref{intro_maps}A provides a logarithmic scale population density distribution map using the data from \textit{D4} and \textit{D5}. The population maps were created as an interpolation of the population density at each antenna.

\section{Collective Mobility Patterns}

We first performed a bulk mobility pattern analysis based on \textit{D2} and \textit{D3} by plotting the probability density function $P(\Delta r)$ of the individual travel distances (or jump sizes) $\Delta r$ in a trace of agglomerated de-identified callers over a period of two weeks, for Ivory Coast (left plot, solid black line) and Portugal (right plot, solid black line), as shown in Figure \ref{fig_commuting}A. The distributions were qualitatively similar to each other except that at the administrative level, the distributions in Ivory Coast are much more scattered than those observed in Portugal, suggesting greater regional variance. We fit the density function to a truncated power law of the form $P(\Delta r) = (\Delta r + \Delta r_0)^{-\beta} exp(-\Delta r/\kappa)$, as described in \cite{gonzalez2008mobility}, where $\Delta r_0$, $\beta$, and $\kappa$ are the fit constants. While the two distributions had similar cutoff distance ($\kappa_{Portugal} = 106 \pm 10$ km; $\kappa_{Ivory} = 122 \pm 5$ km), the two countries have slightly different power law coefficients ($\beta_{Portugal} = 1.37 \pm 0.06$; $\beta_{Ivory} = 1.62 \pm 0.03$). This higher coefficient in Ivory Coast indicates that the likelihood of displacement generally decays faster with distance in comparison to Portugal.

We also investigated regional differences in the mobility patterns. In both cases of Ivory Coast and Portugal, we identified the first level administrative boundaries as the highest country-defined level of partitioning. For Ivory Coast these are called “r\'egions”, while in Portugal they are referred to as “districts”. We partitioned the mobility data by the different level-one administrative regions and overlaid the same density functions specific for each administrative region on the same plots above. Different administrative regions are identified by different scatter marker types and colors. We observed the same truncated power law behavior across the different regions, but the Ivory Coast regions exhibited significantly greater diversity than similarly defined regions in Portugal. This would indicate that in Ivory Coast the likelihood of people migrate and commute with respect to distance is much more dependent on what part of the country they are in, as opposed to in Portugal where the different administrative regions show very little diversity from each other.

Another important metric for assessing mobility patterns is the radius of gyration for the different callers, as defined by the mean squared variance of the center of mass of each user’s set of catchment locations. We computed the radius of gyration using the same method described in \cite{gonzalez2008mobility} and constructed the probability density functions in the same manner as described above; with results shown in Figure \ref{fig_commuting}B we can again observe similar qualitative behaviors. 

The distributions are plotted in Figure \ref{fig_commuting}B and show that the bulk mobility data from Ivory Coast adheres well to the scale-free framework proposed in \cite{gonzalez2008mobility}. The similarity in the bulk mobility characteristics between Ivory Coast and Portugal serves to strengthen the argument that we can make valid comparisons between the two datasets, as described in the sections below.

\section{Commuting Patterns}

Daily commuting patterns are a critical component of any region’s mobility requirements. Displacement is defined as movement from one cell tower to another cell tower between two consecutive calls, and is a key marker for assessing mobility. To focus on daily commuting patterns, we excluded data collected during weekends, and computed the fraction of inter-call events that were accompanied by displacements in a moving 40-minute window of time for Ivory Coast and Portugal. We averaged the fraction of displacement for each 40-minute window across 45 weekdays to get a 24-hour temporal profile of the probability of displacement during a workday. 

\begin{figure}[b!]
\centering
	\includegraphics[height = 17.1cm,trim=2cm 0 2cm 0]{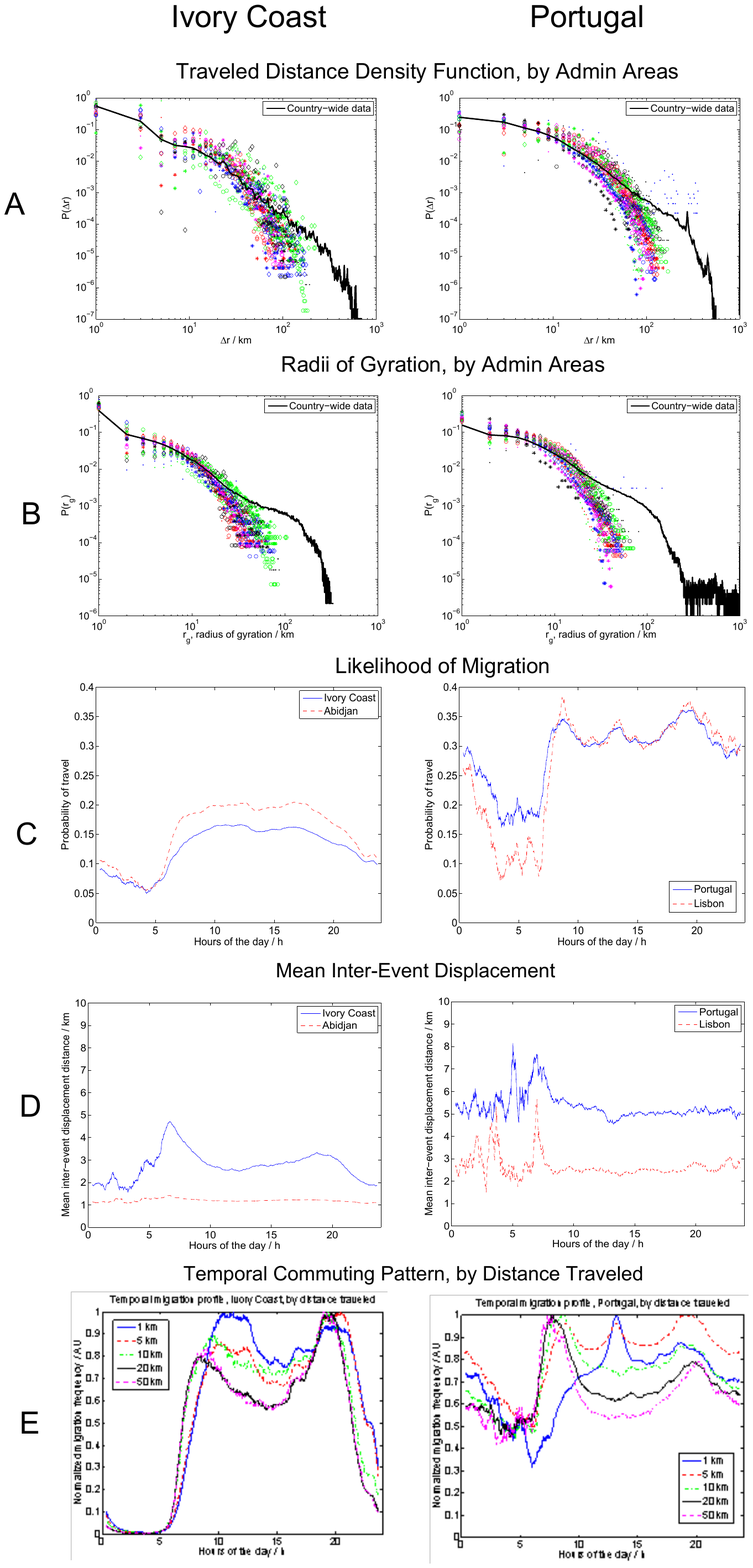}
	\caption{Comparative mobility functions for Ivory Coast (left) and Portugal (right). Probability density functions for  \textit{distance traveled (A)} and \textit{radius of gyration (B)} for each administrative area (differentiated by different color/shape markers). Comparison of country-wide data to capital city daily commuting profiles through respective \textit{probabilities of displacement (C)} and \textit{mean inter-event migration distance (D)}. Country wide temporal commuting profiles \textit{separated by the distance traveled (E)}.}
\label{fig_commuting}
\end{figure}

The first and probably the most significant difference is the absolute difference in the probability of displacement, which can be seen in Figure \ref{fig_commuting}C. We observe that in Portugal, in a given period, people are much more mobile compared to their counterparts in Ivory Coast.

Both countries exhibit a commuting pattern; there is a sharp rise in the probability of displacement around 7-9 a.m. The evening decline is not as sharp, suggesting that people leave work at different times in the evening. 

Significant quantitative differences between the countries can also been seen throughout the day. In Portugal, people in Lisbon and across the nation exhibited similar likelihood to commute during the busiest hours. However, a significantly higher percentage of people in Abidjan were mobile than across the nation. Additionally, while displacement levels in Abidjan and across Ivory Coast were similar during the lowest period (4-7 a.m.). Displacement for the same period is significantly higher for Portugal than for Lisbon, and is likely an indicator of more significant numbers of suburban commuters in Portugal than in Ivory Coast. 

Figure \ref{fig_commuting}D provides a comparison of the mean migration distances between the 2 countries for the same period. Here again, the average distance traveled is significantly less in Ivory Coast and its capital city, than in Portugal. In the country-wide data, we observe a sharp increase in the mean inter-event displacement distance near the morning peak commute (around 5-9 a.m.) in both Ivory Coast and Portugal. However, the spike in distance encountered in Lisbon during morning commute does not occur in Abidjan. This difference may be indicative of people both living and working in close proximity in Abidjan, as opposed to commuting in from outside or across the city as is often the case in developed regions with more comprehensive public transport facilities.

We examined the country-specific commuting pattern more closely by looking at how the distance commuted may affect the daily behaviors. The observed distances traveled were binned (0-1 km, 1-5 km, 5-10 km, 10-20 km, and 20-50 km), and the daily temporal profile of the probability of displacement was computed for each bin for the two countries, as shown in Figure \ref{fig_commuting}E.

By looking at the figures closely, we can draw many interesting and relevant insights regarding commuting. Firstly, for Ivory Coast (left plot), note that regardless of the distances commuted, the temporal profiles show a bimodal pattern: a morning peak (around 9 a.m.) as well as an evening peak (around 7 p.m.), as expected from a typical commuting pattern. Between the two peaks, there is a valley which deepens as the distances commuted increases. This also makes intuitive sense, as people who live far away from their work place are likely only going to make the long commute twice a day, and there are only rare occasions during the work day where such a migration is required. Whereas, for shorter distances traveled (such as the solid blue line or the dashed red line), this valley is less prominent. This difference likely is explained by the fact that during the day, people at work may make frequent short trips, such as to visit their clients, to replenish their inventories, etc.

Comparing this pattern in Ivory Coast with that in Portugal (right plot), we observe further differences. First, while the bimodal behavior is largely preserved in Portugal, it is less prominent for shorter commuting distances (in particular, in the 0-1 km bin, as shown in the solid blue line). In fact, in bins associated with shorter displacement distances, a third peak occurs around midday (13:10 for the 0-1 km bin and 13:09 for the 1-5 km bin, respectively), which is absent in Ivory Coast and in the bins with longer distances traveled in the Portugal data. This third peak likely represents people going for lunch, and it is only observed in short-distance displacements, as it is rare for people to travel for long distances for lunch when they are at work. The absence of this feature in the Ivory Coast data can be accounted for by the fact that in Portugal, workers are required not allowed to work for more than five consecutive hours and are subsequently forced to a break for at least one hour in the middle of the workday, which gives them more chance to travel further for lunch or other reasons \cite{portugal_worktimes}.

\section{Community Structure}

Large networks, such as the telecommunications or transportation networks of a nation, often exhibit community structure, i.e., the organization of vertices into clusters with many edges joining vertices of the same cluster and comparatively few edges joining vertices of different clusters. Identifying the community structure in such networks has many applications, such as better placement and provisioning of services. Recently, this type of community structure analysis has been performed on landline communications in Great Britain \cite{ratti2010uk}, mobile connections in Belgium \cite{expert2011uncovering}, United States \cite{calabrese2011connected}, and various other countries across Europe, Asia, and Africa \cite{sobolevsky2013delineating}. While research done to investigate the impact that physical human mobility has on the space-independent community structure has been also explored in \cite{calabrese2011connected} there has been a lack of similar research, especially  in the developing context. 

Network modularity \cite{newman2006modularity} is a measure of the strength of the division of a network into clusters. Networks with high modularity have dense connections between nodes within clusters, and sparse connections between nodes in different clusters. Modularity is computed as the fraction of edges that fall within a cluster, minus the expected such fraction if the edges were distributed at random with respect to the node strength distribution. The value of modularity lies in the range [-1 1], and is positive if the edges within groups exceeds the number expected on the basis of chance.

A rigorous definition of network modularity is:

\begin{equation} 
  Q=\frac{1}{2m}\sum_{ij}\left [ A_{ij} - \frac{ k_{i}k_{j} } {2m}  \right ] \delta( c_{i}, c_{j} )
\end{equation}

where $A_{ij}$ is the weight of the link from $i$ to $j$, $k_{i}$ is the sum of the weights from node $i$, $c_{i}$ is the community that node $i$ was assigned to, $m=\frac{1}{2}\sum_{ij} A_{ij}$, and $\delta( c_{i}, c_{j})$ is 1 if $c_{i}=c_{j}$ and 0 otherwise.

High modularity in mobility networks may point to an efficient organization of residences, employment, and services all in close proximity, or it may point to restrictive policies or infrastructures that limit free movement across communities. We were interested in the community structure of developing nations, such as Ivory Coast, especially in comparison to more developed nations, such as Portugal. 

We used datasets \textit{D2} and \textit{D3} to build the human mobility networks and identify the community structure of antennae within the Ivory Coast and Portugal. We set nodes to the locations of each cell tower, and edges to the total number of migrations of all people that placed two consecutive calls between the two nodes within a time window of 24 hours. We tested the following Community Detection algorithms: Louvain \cite{blondel2008louvain}, Le Martelot \cite{lemartelot2011}, Newman \cite{newman2006}, Infomap \cite{lancichinetti2009community}, and a new method of community detection suggested in \cite{sobolevsky2013combo}. We computed the modularity of community structures identified by each of these methods. The method described in \cite{sobolevsky2013combo} provided the highest modularity, and was subsequently chosen to be used for this part of the study. Figure \ref{intro_maps}C graphically compares the communities identified (in color) with their first level administrative boundaries (outlined in black). 

An especially interesting difference in the communities identified for Ivory Coast and those identified for Portugal was the similarity between identified communities and the official administrative boundaries of the nations. We calculated 7 different clustering coefficients, each representing the qualitative similarity between the two different partitions of each region. While the communities identified for Portugal exhibited high similarity with the 20 official administrative boundaries (districts), this was not the case for the Ivory Coast's 19 official administrative boundaries (r\'{e}gions). As shown in Table 1, communities identified for Portugal show significantly higher similarity (as much as 28\% higher clustering coefficients) to administrative boundaries, in comparison to that of the Ivory Coast.

\begin{table}[h!]
\centering
    \begin{tabular}{|l||l l l| p{1cm}} \hline
    \textbf{Similarity Index}   				& \textbf{Portugal} & 	\textbf{Ivory Coast} & \textbf{Difference} \\ \hline
    Wallace \cite{wallace1983} 				& 0.482813 &		0.199134 &		0.283679 \\			
    Adjusted Rand \cite{rand1971clustering} 		& 0.495536 &		0.258351 &		0.237184 \\
    Jaccard \cite{jaccard1901etude} 			& 0.377999 &		0.184733 &		0.193266 \\			
    Fowlkes-Mallows \cite{fowlkes1983clustering} 	& 0.553788 &		0.378296 &		0.175491 \\			
    Melia-Heckerman \cite{heckerman1998clustering} 	& 0.659385 &		0.515347 &		0.144037 \\			
    Hubert \cite{hubert1985comparing} 			& 0.806411 &		0.707795 &		0.098615 \\			
    Larsen \cite{larsen1999} 				& 0.58273 &			0.525396 &		0.057334 \\			
    Rand \cite{rand1971clustering} 			& 0.903205 &		0.853898 &		0.049308 \\			\hline
    \end{tabular}-
\caption{\textmd{Comparison of different similarity indices to compare similarity between community partitioning (generated from network of human mobility) and the respective administrative boundaries.}}
\end{table}

While this significant difference in community and official boundary alignments may be attributable to the layout of infrastructure along official boundaries, we began to question whether there might be more fundamental differences. Previous studies have shown that other factors, such as geographical features, can play an important role in how communities are formed and services are sought \cite{ratti2010uk,onnela2011geographic}. However, little has been done to investigate the direct impact of culture and language on human mobility.

Ivory Coast represents an especially interesting context to investigate cultural and linguistic influences on mobility within a single nation. The Ivory Coast is a nation made up of more than 60 distinct tribes, classified into 5 principle regions \cite{ic_online}. The official language is French, although many of the local languages are widely used, including Baoulr\'{e}, Dioula, Dan, Anyin and Cebaara Senufo, and an estimated 65 languages are spoken in the country. 

Intuitively, these cultural and linguistic differences are likely to influence mobility patterns in the region. However, it is also known that as regions become more urbanized, cultural ways are often blended or lost altogether. Portugal represents an interesting context for the latter, as Portuguese is the single national language of Portugal, and any tribal boundaries pre-date Roman times.

Due to the vast differences seen in the network community structure to administrative-defined boundaries, our goal in the next section was to understand if tribal structure of the Ivory Coast could be the influential factor of the mobility patterns in the region.

\section{Tribal Community Analysis}

\begin{figure}[h!]
\centering
	\includegraphics[width=8cm,height = 9cm]{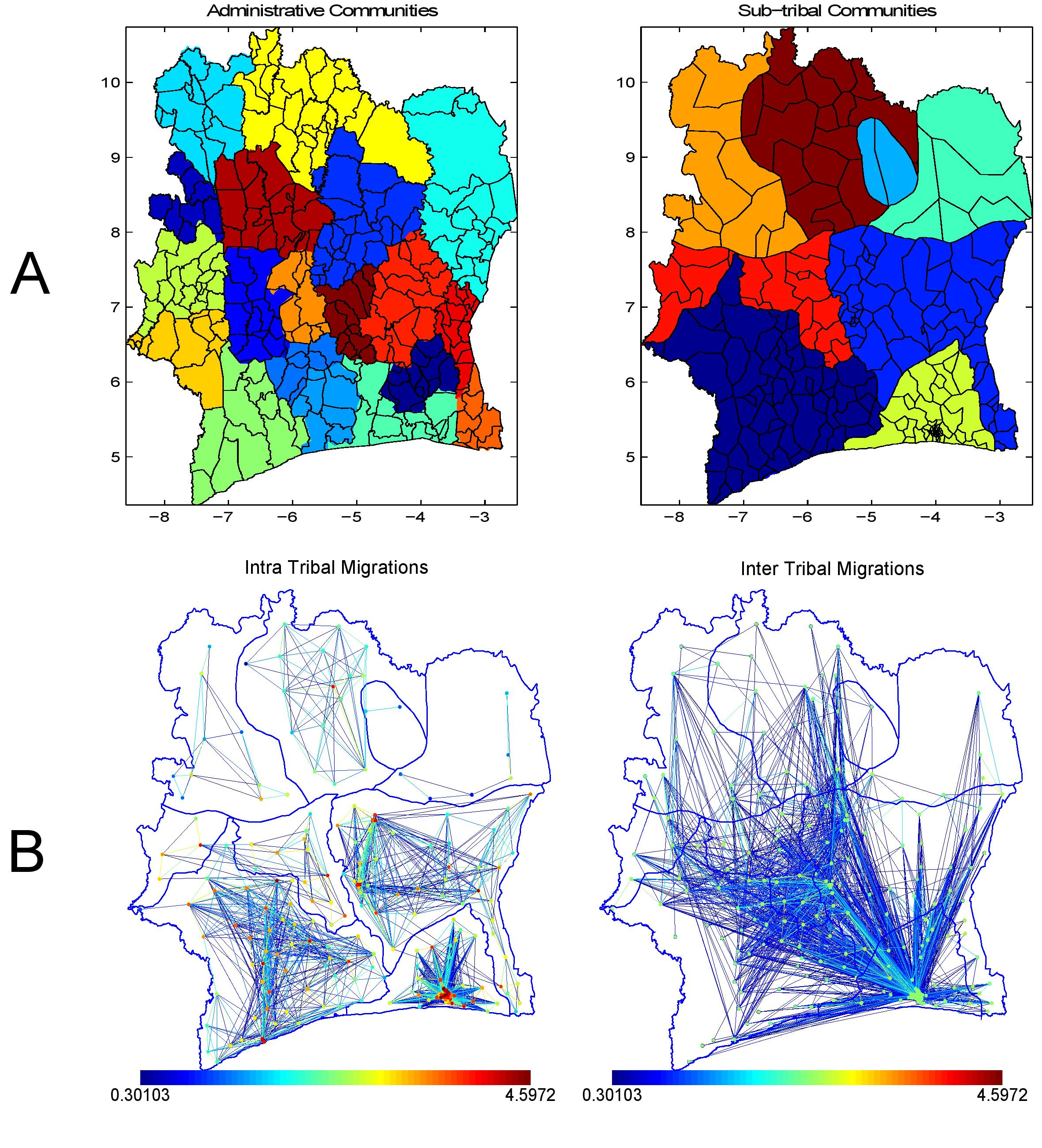}
	\caption{\textit{A. Partitionings of Ivory Coast} by administrative prefectures/sub-prefectures (left) and tribal/sub-tribal communities (right). \textit{B. Intra-Inter tribal migrations}, where each node represents an individual sub-tribal community, and each link is logarithmically colored to represent the number of migrations (extracted from call records) between the two nodes.}
	\label{fig_ic_maps}
\end{figure}

Since the communities detected in Section IV exhibited low similarity to administrative boundaries, we began to investigate the impact of certain factors present in the Ivory Coast that may be attributable to such a substantial difference. Since people and their behaviors throughout a large portion of Africa are still impacted by their tribal affiliations while any tribal boundaries in Portugal predate Roman times, studying the tribal boundary impact in Ivory Coast represented a perfect example of this. We started by modifying the community detection approach to use the tribal boundaries as the Level 1 boundaries and subsequently ran a hierarchical community detection using the Louvain method \cite{blondel2008louvain} in each of these Level 1 tribal partitions in order to produce the subcommunities inside different tribal regions which we will refer to as sub-tribal communities. The Louvain method provided the closest final number of partitions to the administratively defined subprefectures of the algorithms that we tested, and was thus chosen as the most appropriate for current purposes. By doing so, we were able to generate sub-tribal communities while also conserving the physical shape of each tribal region. Figure \ref{fig_ic_maps}A demonstrates this by showing the official prefecture and subprefecture boundaries (left) and the tribal and sub-tribal communities (right) that were created using this approach. The larger first level partitions (prefecture and tribal) are indicated as a single color, while the smaller secondary partitions (subprefecture and sub-tribal) are indicated as black lines within their respective first level partition.

As a first measure of impact of tribes on mobility, we aggregated the mobility network between communities. Figure \ref{fig_ic_maps}B provides a plot of the mobility network with each node representing a community, and each edge colored to reflect the number of migrations between the connected nodes in the mobility network. The intra-tribal community mobility network is plotted separately from the inter-tribal community mobility network, in order to facilitate comparison. The intra-tribal network plots only those edges between communities in the same tribe. The inter-tribal network plots only edges between communities of differing tribes. 

This diagram provides a first insight into tribal influences on mobility. Note that the number of intra-tribal migrations (as indicated by the color coding of edges) dwarfs the number of inter-tribal migrations. Additionally, the inter-tribal migrations are largely dominated by connections to the largest city, Abidjan. 

\subsection{Modelling Human Mobility}

Accurately modeling human interactions between regions can present many challenges; however, effectively doing so can provide a crucial piece of information to efficiently distribute resources, health services, etc. throughout a given area. The Gravity Model, whose origins trace back to Ravenstein's laws of migration \cite{ravenstein1885laws}, was formulated on Newton’s Law of gravity, and predicts flux between a source and destination based on the populations of the source and destination, and the distance between the source and destination. More specifically, according to the gravity model, the average flux migrations from regions $i$ to $j$ is: 

\begin{equation} 
  T_{ij}=\frac{m_{i}^{\alpha}n_{j}^{\beta}}{r_{ij}^{\gamma}}
\end{equation}

where $i$ and $j$ are origin and destination locations with populations $m_i$ and $n_j$ respectively at a distance of $r_{ij}$ from each other. $\alpha$, $\beta$, $\gamma$ are adjustable parameters chosen to fit the data.

In order to apply the model one must inherently define the appropriate spatial resolution (ie. the areas to model migrations between/within). By monitoring the resulting accuracy of the model, it is possible to gain insight on what type of partitioning of an area will most effectively allow for human mobility to be modeled. We started by investigating if the use of these sub-tribal communities would provide an advantage in modeling the mobility network of the Ivory Coast compared to an administrative (subprefecture) partitioning. 

We computed the Gravity model using dataset \textit{D2}, and specifically modeled the migrations for both administrative and sub-tribal partitioning of the country, and subsequently tested the accuracy of the model from the mean average percent error (MAPE) with respect to the true network of human mobility. MAPE has been shown to be a very effective measure of error in model predictions, especially when considering population forecasting \cite{swanson2011mape, tayman1999validity}.

We compute Mean Average Percent Error (MAPE) according to:

\begin{equation} 
  M=\frac{100}{n} \sum_{t=1}^{n}\left | \frac{A_t-F_t}{A_t} \right |
\end{equation}

where $A_t$ is the actual value, $F_t$ is the forecasted value, and $n$ is the number of data points. Therefore, an inaccurate model will subsequently yield a high MAPE value; whereas an accurate model will yield a much smaller MAPE value.

In order to further explore the relationship between tribal and administrative boundaries we applied an alternative approach for modeling human mobility and interaction, the Radiaiton Model. The Radiation Model \cite{simini2012radiation} was recently proposed as a parameter free mobility model in which individuals move and interact based on the population density of the source and destination regions, and that of the surrounding regions. Using the Radiation Model, the average flux between two regions $i$ and $j$ is:

\begin{equation} 
  T_{ij}=T_{i}\frac{m_{i}n_{j}} {(m_i+s_{ij})(m_i+n_j+s_{ij})}
\end{equation}

where $i$ and $j$ are origin and destination locations with populations’ $m_i$ and $n_j$ respectively, at distance $r_{ij}$ from each other, with $s_{ij}$ representing the total population in the circle of radius $r_{ij}$ centered at $i$ (excluding the source and destination population). $T_i$ signifies the total outgoing flux that originates from region $i$.

\begin{figure}[h!]
	\begin{center}
		\includegraphics[width=0.9\textwidth]{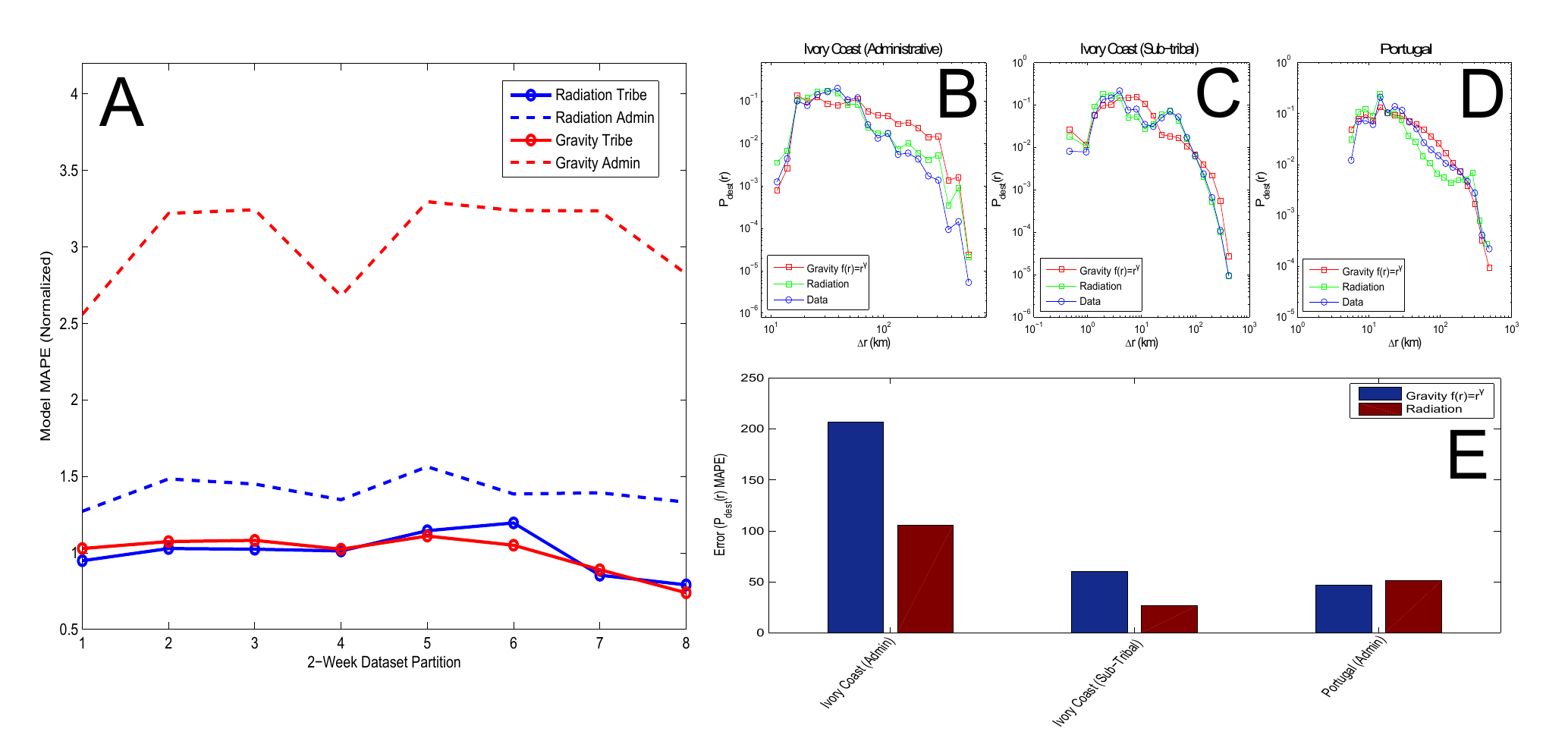}
	\end{center}   
	\caption{\textit{A. Normalized MAPE values} for Administrative and Tribal paritionings for Radiation and Gravity Model. The migration models for \textit{Ivory Coast’s administrative (B)} and \textit{sub-tribal boundaries (C)}, as well as in \textit{Portugal’s administrative partitioning (D)} are also compared where each signal represents the probability of migration to a location ‘r’ kilometers away from the originating location. MAPE values for each model and partitioning to the respective data are shown in \textit{E}.}
\label{fig_admin-v-tribe}
\end{figure}

Figure \ref{fig_admin-v-tribe}A shows the normalized MAPE comparison for the sub-tribal and administrative boundaries, and demonstrates that the MAPE for Gravity Model predictions made via administrative boundaries ranged from 150\% to 230\% higher than that of sub-tribal communities, while the Radiation Model also yielded 20\% to 50\% higher MAPE values for administrative boundaries. This indicates a higher accuracy (lower MAPE) produced when sub-tribal communities were used, as opposed to administrative boundaries. This subsequently suggests that, in terms of mobility patterns of Ivory Coast, it is more efficient to model mobility on a partition that accounts for tribal, cultural, and lingual differences in groups of people, as opposed to the current administratively defined country partition.

We also partitioned the mobility model predictions according to intra-tribal and inter-tribal flux in order to quantify the \textit{strength of the connectivity of the tribes}. Quantitatively, the MAPE for the inter-tribal mobility was 11.3\% higher than the MAPE of the intra-tribal mobility predictions, and supports the dominant pattern of intra-tribal migrations over inter-tribal migration which may require special consideration in terms of mobility modeling. Again, the fact that the Radiation model produces more accurate results for migrations within a single tribe compared to those between tribes suggests that the tribes themselves are playing a key role in the overall improved accuracy of the model. 

Figures \ref{fig_admin-v-tribe}B-D compare the predicted probability of migration versus the ground truth migration, for both the Gravity Model and Radiation Model. As a more direct comparison of accuracy, Figure \ref{fig_admin-v-tribe}E provides the error (MAPE) for the models plotted in Figures \ref{fig_admin-v-tribe}B-D. For the Ivory Coast, these figures show the higher accuracy (i.e., lower error) of Radiation Model for both administrative and sub-tribal communities, and it shows that using the Radiation Model with sub-tribal communities provides the highest accuracy (i.e., lowest MAPE). 

Figure \ref{fig_admin-v-tribe}D illustrates the Portuguese administrative municipality boundaries perform well for both the Radiation and Gravity Models. This may be indicative that the municipal boundaries were designed to align with cultural and social communities or that cultural and social communities have adapted to fit administrative boundaries. 

However, we believe a more likely explanation is the growing homogeneity of language and culture that comes with maturing industrialization and urbanization. This is reflected in the predominance of Portuguese as the national language in Portugal, compared to the more than 60 local languages spoken in Ivory Coast.

There are several important implications from these findings.
\begin{enumerate}
  \item Models of mobility, migration, and interaction that are conceptualized in mature, industrialized, and urbanized regions may not directly map to developing regions with more pronounced cultural and linguistic differences. Such models need to better account for these differences.
  \item If administrative boundaries are drawn and services are placed based on models that do not accurately reflect these influencers, results could include inefficiencies, leading to inequality of services (e.g., longer or less accessible commutes), and potentially discrimination and alienation of segments of the population.
  \item Techniques for assessing the strength of borders in a given regional partitioning scheme are critical to ensuring the accuracy of mobility and migration modeling, and, perhaps more importantly, to enabling sound decision-making by authorities tasked with setting effective administrative boundaries.

\end{enumerate}

In the following section, we propose novel techniques for quantifying the impact of a regional partition scheme on model accuracy, and for assessing the strength of borders in a given partitioning scheme.

\section{Assessing regional affinities and border strength}

In previous sections, we demonstrated the issues arising from using mobility models, such as the Gravity and Radiation models, on regions where the partitioning scheme, such as Ivory Coast administrative boundaries, does not reflect the regional affinities and border strengths these models assume. We illustrated techniques to create more appropriate partitioning schemes, such as the tribal communities, and demonstrated the ability to achieve higher accuracy mobility modeling using this improved partitioning. 

However, it may not always be possible to simply re-draw borders. Instead, tools are needed to assess the efficacy of an existing partitioning scheme, in terms of the affinities within the identified borders and the strength of the borders. 

In this section, we propose two novel techniques to address the above challenge. Firstly, we present a metric to determine whether affinities exist within borders that may impact the accuracy of mobility modeling. We test our metric on the tribal and administrative boundaries used in the previous sections. Secondly, we propose a technique to assess the strength of existing borders, and we demonstrate our technique on the existing administrative borders of Portugal and Ivory Coast. Our techniques use the same mobility data used in previous sections and can be performed on any regional partitioning scheme, and thus provide valuable tools to mobility researchers and to urban planners.

\subsection{Regional affinity}

 The accuracy of both the Gravity Model and the Radiation Model depends on the ability to accurately model, for a given time epoch, movement from any region to any other region, and lack of movement to another region. Inter-region movements are driven by opportunities and resources, which are reflected in the population of region, and constrained by distance. Intra-region affinities, such as physical proximity to home, work, family, and friends, tend to limit movement from a given region. Improperly partitioning a region to account these affinities results in over or under predicting flux, and therefore, poor model performance. 

We propose a metric to assess whether such affinities exist, and test that metric on the tribal and administrative boundaries used in previous sections. To compute this metric, we segregated all migrations into two categories: intra- and inter- regional migrations. Referring to Figure \ref{fig_ic_maps}A, an inter-region migration is a migration that crosses a color boundary. Likewise, an intra-region migration is a migration that does not cross a color boundary. When a model over-predicts the number of inter-region migrations, it is under-estimating the strength of the affinities within that region. 

We use $S$ to denote the bias of a regional partitioning to over or under estimate flux across regional boundaries. We compute $S$ as:

\begin{equation}
   S=\frac{ \sum_{1}^{n} \frac{R}{T} }{n}
\end{equation}

where $R$ is the modeled flux, $T$ is the true flux, and $n$ is the total number of predictions made. In other words, $S$ represents the average ratio of modeled to predicted fluxes in the system. Therefore, relatively high values of $S$ indicate the model is generally \textit{over} predicting, while lower values indicate a general \textit{under} prediction by the model. We compute separate $S_{inter}$ and $S_{intra}$ values, where the $S_{inter}$ value includes only inter-region flux and $S_{intra}$ includes only intra-region flux. 

\begin{figure}[h!]
	\begin{center}
		\includegraphics[keepaspectratio=true, height=6cm]{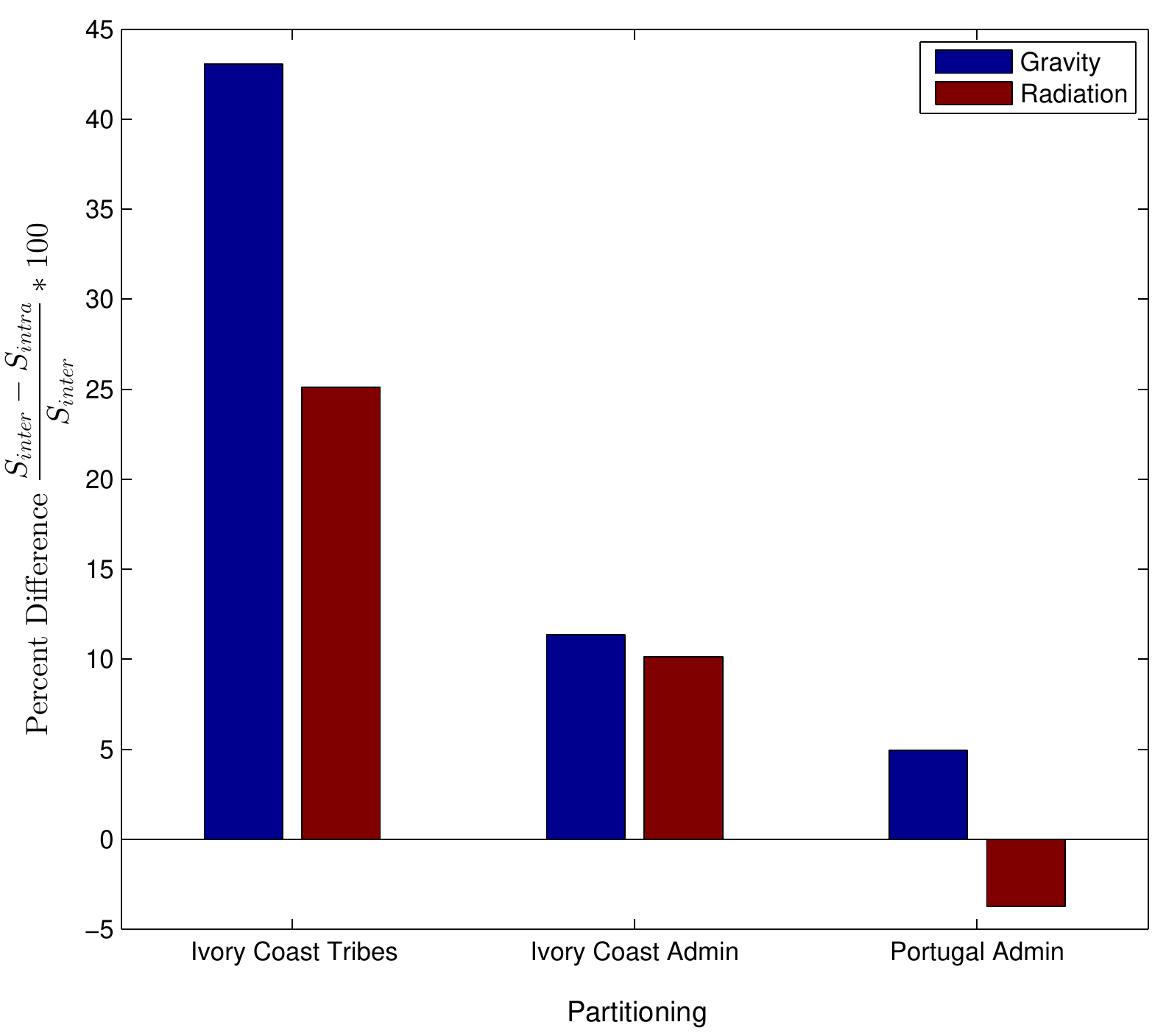}
	\end{center}   
	\caption{Percent difference in values of $S$ for intra and inter migration bias towards holistic model accuracy.}
	\label{model_slope}
\end{figure}

The percent difference ($D$) between $S_{intra}$ and $S_{inter}$ for all partitions provides a measure of bias. That is, the larger the value of $D$, the stronger the bias for over estimating inter-region flux. Figure \ref{model_slope} illustrates that both the Gravity and Radiation Models exhibited the most significant over estimation of inter region flux for Ivory Coast Tribes. Recall that significant over estimation of inter region flux is a result of significant under estimation of intra region affinities. Consistent with the results of previous sections, the administrative boundaries of both Portugal and Ivory Coast demonstrate substantially less bias. Thus our metric performs well in highlighting the strong affinities (in this case tribal), which must be accounted for in the model.

\subsection{Border Strength}

Just as mobility may be constrained by affinities, it is similarly impacted by the strength of surrounding borders. Mobility modelling must accurately and compactly reflect borders that may be physical, such as gates or other guards, or abstract, such as lack of opportunities. 

In this section, we propose a novel metric for assessing the strength of borders within a region. We demonstrate our metric by computing the strengths of the administrative borders of Portugal and Ivory Coast used in previous sections. As predicted by the Gravity and Radiation Models, we show that borders surrounding the heavily populated region of Abidjan are more penetrable than borders elsewhere in the Ivory Coast. Similarly, our results show that the borders throughout the country of Portugal are much more uniform, even showing consistency with the more heavily populated Lisbon. Our metric allows us to show that while these two cities, located in two very different regions, share a higher penetrability of borders than surrounding regions, they also exhibit significant differences in the distribution of border strength. 

To compute our border strength metric, we start by computing the connectedness of each node (in our study, cell-tower) to the partition of which it is a member. We use $P$ to denote the partitions to be evaluated, $p$ as the total number of partitions, and $P_i$ to denote the partition in which $i$ is a member. We compute the connectedness of each node to each partition as $C$, where:

\begin{subequations}
  \begin{equation}
    e_{i,j}= m_{i,j} -  S_i  T_j
  \end{equation}
  \begin{equation}
    C_{i,P}=  \frac{ \sum_{\substack{j \in P\\j \ne i }}{\left(   e_{i,j} + e_{j,i}   \right)}  } {S_i+T_i-2m_{i,i}}
  \end{equation}

where $i$ and $j$ are nodes, $w_{i,j}$ is the weighted directed matrix of human mobility, $m_{i,j}=\frac{w_{i,j}}{\sum{w_{i,j}}}$, $S_i=\sum_j{m_{i,j}}$, and $T_j=\sum_i{m_{i,j}}$. Thus for any given node $i$, there exists a partition $P_q$, such that $C_{i,P_q}$ is greater to or equal to all other values of $C_{i,P}$. Also, if $P_q=P_i$, that is, the partition with the greatest connectedness is $P_i$, then node $i$ belongs to the partition to which it is most strongly connected. $e_{i,j}$ is computed as the difference between the actual number of migrations and the expected value of migrations from $i$ to $j$.

 We use the connectedness, $C$, of each node to compute a measure of how well connected a node is to it's assigned partition as $s_i$ where:

  \begin{equation}
    s_{i}=C_{i,P_i} - \max_{D \ne P_i}{\{C_{i,D}\}}
  \end{equation}
  \label{strength_equations}
\end{subequations}

Therefore, for any given node $i$, $-1 \leq s_i \leq 1$, where $s_i=0$ indicates node $i$ is as strongly connected to at least one partition other than the partition $p-i$ to which it is assigned. Negative values of $s_i$ indicate node $i$ is more strongly connected to partitions other than the partition $P_i$ to which it is assigned, and positive values of $s_i$ indicate node $i$ is more strongly connected to it's assigned partition $P_i$.

\begin{figure}[b!]
	\begin{center}
		\includegraphics[keepaspectratio=true, width=0.97\textwidth]{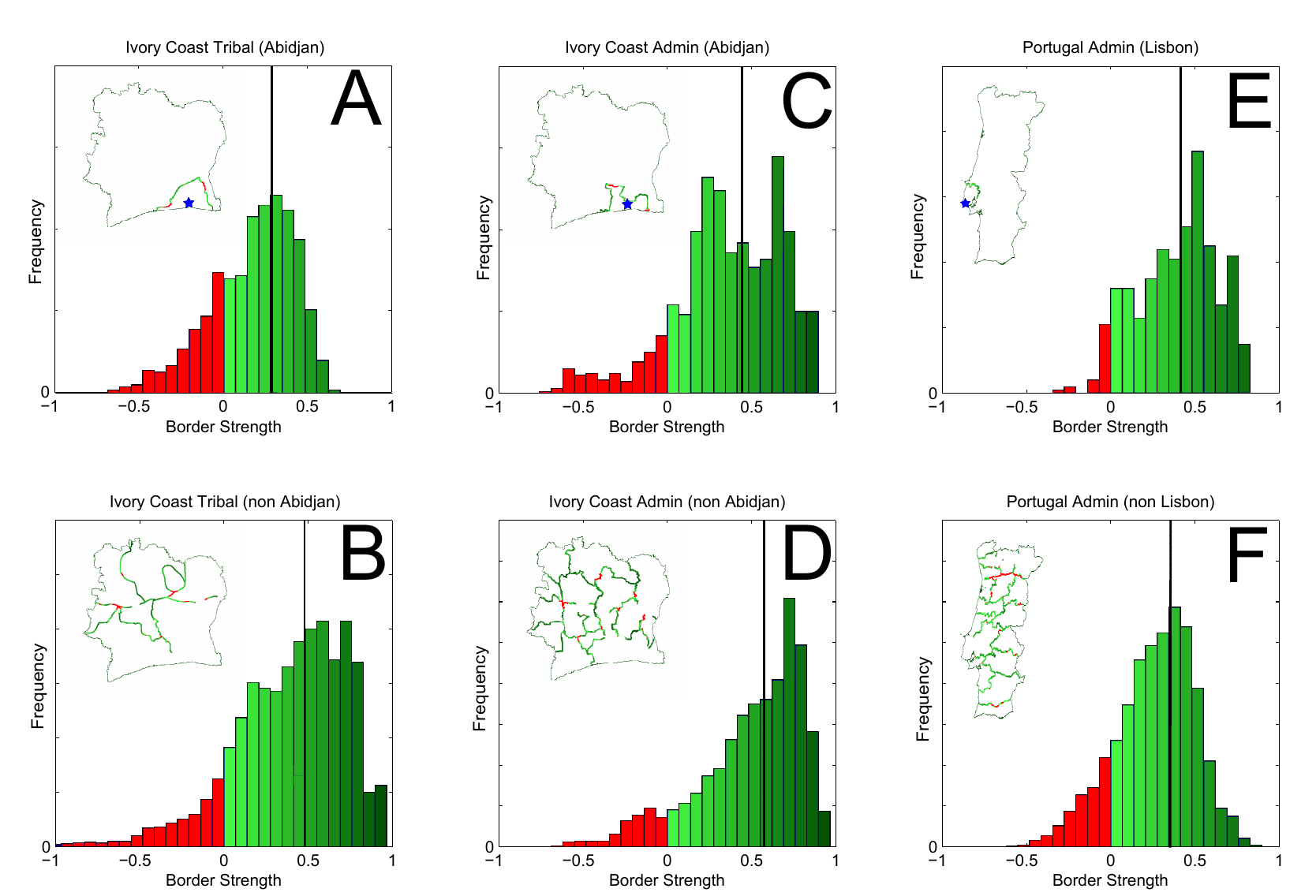}
	\end{center}   
	\caption{Strength of node connectivity visually mapped from values of $s_i$ along \textit{A-B Ivory Coast Tribal}; \textit{C-D Ivory Coast Administrative}; and \textit{E-F Portuguese Administrative} boundaries. Border categories are separated according to whether they contain the main city of the respective country (ie. \textit{A \& C. Abidjan, Ivory Coast} and \textit{E. Lisbon, Portugal}). Histograms display the distribution of the border strength values along the border(s). Black vertical lines are indicative of the mean positive border strength value.}
\label{hist_border}
\end{figure}

We overlay the $s_i$ values onto the corresponding map of the region and perform a linear interpolation of these values in order to assess connectedness across the borders. Figure \ref{hist_border} illustrates the evaluation of this connectedness value along the different borders of Ivory Coast and Portugal. Furthermore, calculating the mean of positive $s_i$ values for only points that lie on the border of interest creates a generalized measure of the strength of the borders in the region ($\overline{s_i}$). Figures \ref{hist_border}A-\ref{hist_border}D illustrate how significantly more penetrable borders surrounding the densely populated Abidjan are in comparison to other portions of the country. This higher penetrability of borders surrounding Abidjan holds for both tribal (Figure \ref{hist_border}A) in comparison to the remainder of tribal borders (Figure \ref{hist_border}B). This effect is again prominent with the administrative  border surrounding Abidjan (Figure \ref{hist_border}C) relative to the rest of the country (Figure \ref{hist_border}D). Furthermore, Figure \ref{hist_border}A illustrates the mean positive penetrability ($\overline{s_i}$) of 0.2847 for the borders surrounding Abidjan, versus 0.4767 for borders throughout the rest of Ivory Coast (Figure \ref{hist_border}B). Also note that penetrability in Abidjan is markedly higher than in the remainder of Ivory Coast, whereas the Lisbon borders (Figures \ref{hist_border}E-\ref{hist_border}F) are only marginally less penetrable than in the remainder of Portugal. This supports our earlier finding that mobility in Portugal is much more uniform throughout the entire country compared to the Ivorian Coast context.

The distributions illustrated in Figure \ref{hist_border} also provide interesting insights into mobility in Ivory Coast and Portugal. More specifically, notice the tight clustering of border strengths for Abidjan under the Tribal partitioning (Figure \ref{hist_border}A) versus more widely distributed border strength values for the rest of Ivory Coast. This supports our early findings that Tribal boundaries play a key role in mobility throughout the country, except in highly urbanized regions such as Abidjan.

\section{Conclusion}

Africa is a continent that has been shaped by human migration over tens of thousands of years. Indeed, migrations within and beyond the African borders have recently been shown as influencing all civilizations as we know them. However, until recently, there has been a dearth of data on the forms and patterns of migration within the nations of Africa. Moreover, much of the mobility research is based on theories that have emerged from highly industrialized nations and lack validation in the context of developing environments.

Our study has demonstrated that many of these conceptions are not necessarily applicable in the African context. We have made these differences clear by comparing our findings in Ivory Coast to one such industrialized nation, Portugal. For example, we have shown that the probability of displacement during normal commuting hours in Portugal is often nearly double that of Ivory Coast for the same time of day. Similarly, average distances traveled by commuters in Portugal is nearly double that of commuters in Ivory Coast.

While differences in the likelihood of travel and average distance travel can be attributed to quantitative differences in infrastructural support for mobility this already strongly affects the whole mobility picture leading to a number of quantitative dissimilarities. Our study shows evidence of more fundamental differences in infrastructural support for mobility, such as tribal, cultural, and lingual differences. In addition, we demonstrate that the similarity between administrative boundaries and communities detected in mobile phone data is markedly lower in Ivory Coast than in Portugal. 

By identifying the tribal influence on mobility in the Ivory Coast, we were able to illuminate further differences in mobility patterns. For example, we were able to show intra-tribal migrations were much more frequent than that of inter-tribal migrations over the same distance and therefore are under or overestimated by the models. Taking this into account by exploiting our tribally aligned communities for the mobility models drastically improves modeling of human mobility in Ivory Coast. We validated this higher accuracy by computing the Mean Absolute Percentage Error (MAPE) across all data points for both models, and found a 20\% to 50\% higher error for the models using administrative boundaries. We also validated our results by computing the distribution of migrations by distance migrated and found that by using this sub-tribal method of spatial units definition in modeling human mobility we were able to improve the accuracy of the models so drastically, that the Ivory Coast performed even better than its developed country-counterpart, Portugal.

We propose novel techniques for assessing the strength of borders within a regional partitioning scheme, and for assessing the impact of inappropriate partitioning on model accuracy. Our results offer improved insights on why models developed for mature and stable regions may not translate well to developing regions, and provide tools for urban planners and data scientists to address these deficiencies. 

We are excited by the findings of this study and plan to further validate our findings by comparing to additional developing and developed regions.

\section{Acknowledgements}
The authors wish to thank Orange and D4D Challenge for providing the datasets used throughout this study. We further thank Ericsson, the MIT SMART Program, the Center for Complex Engineering Systems (CCES) at KACST and MIT CCES program, the National Science Foundation, the MIT Portugal Program, the AT\&T Foundation, Audi Volkswagen, BBVA, The Coca Cola Company, Expo 2015, Ferrovial, The Regional Municipality of Wood Buffalo and all the members of the MIT Senseable City Lab Consortium for supporting the research.

\nocite{stanbookchapter}

\bibliographystyle{bmc-mathphys} 
\bibliography{refrences}      

\end{document}